\documentclass[a4paper,11pt]{article}
\pdfoutput=1 

\usepackage{jheppub} 

\usepackage[T1]{fontenc} 

\usepackage{dsfont}

\def\ben{\begin{equation}}
\def\een{\end{equation}}
\def\bea{\begin{eqnarray}}
\def\eea{\end{eqnarray}}

\title{Three Dimensional View of Arbitrary $q$ SYK models}

\author{Sumit R. Das$^1$,}
\author{Animik Ghosh$^1$,}
\author{Antal Jevicki$^2$,}
\author{Kenta Suzuki$^2$}

\affiliation{$^1$Department of Physics and Astronomy, University of Kentucky, Lexington, KY 40506, U.S.A.}
\affiliation{$^2$Department of Physics, Brown University, 182 Hope Street, Providence, RI 02912, U.S.A.}

\emailAdd{das@pa.uky.edu}\emailAdd{animik.ghosh@uky.edu}\emailAdd{antal\_jevicki@brown.edu}\emailAdd{kenta\_suzuki@brown.edu}

\abstract{In \url{arXiv:1704.07208} it was shown that the spectrum and bilocal propagator of SYK model with four fermion interactions can be 
realized as a three dimensional model  in $AdS_2 \times S^1/Z_2$ with nontrivial boundary conditions in the additional dimension. In this paper we show that  a similar picture holds for generalizations of the SYK model with $q$-fermion interactions. The 3D realization is now given on  a space whose metric is conformal to $AdS_2 \times S^1/Z_2$ and is subject to a non-trivial potential in addition to a delta function at the center of the interval. It is shown that a Horava-Witten compactification reproduces the exact SYK spectrum and a non-standard propagator between points which lie at the center of the interval exactly agrees with the bilocal propagator. As $q \rightarrow \infty$, the wave function of one of the modes at the center of the interval vanish as $1/q$, while the others vanish as $1/q^2$, in a way consistent with the fact that in the SYK model only one of the modes contributes to the bilocal propagator in this limit.}

\begin{document}

\begin{flushright}
{UK/17-10} \\ {BROWN-HET-1727}
\end{flushright}

\maketitle
\flushbottom

\section{Introduction}
\label{sec:intro}
The Sachdev-Ye-Kitaev (SYK) model \cite{Sachdev:1992fk, Sachdev:2010um, Sachdev:2015efa, Kitaev:2014, Kitaev:2015} has in recent intensive studies  \cite{Polchinski:2016xgd, Maldacena:2016hyu, Jevicki:2016bwu, Jevicki:2016ito, Davison:2016ngz, Stanford:2017thb,Mertens:2017mtv, Gross:2017hcz, Gross:2017aos, Kitaev:2017awl, Das:2017pif}
(for a review see \cite{Sarosi:2017ykf}), emerged as a useful toy model for holography. An important aspect of this model, which is absent from other such models of holography is the presence of quantum chaos \cite{Shenker:2013pqa, Leichenauer:2014nxa, Shenker:2014cwa, Maldacena:2015waa, Polchinski:2015cea, Caputa:2016tgt, Gu:2016hoy, Perlmutter:2016pkf, Anninos:2016szt, Turiaci:2016cvo}, which indicates that a finite temperature version of this model describes black holes. Related models have been studied \cite{Danshita:2016xbo, Erdmenger:2015xpq, Krishnan:2017lra}
with extensions \cite{Gross:2016kjj, Gu:2016oyy, Berkooz:2016cvq, Fu:2016vas, Fu:2016yrv, Garcia-Alvarez:2016wem, Hartnoll:2016mdv, Blake:2016jnn, Nishinaka:2016nxg, Turiaci:2017zwd, Jian:2017unn, Chew:2017xuo, Yoon:2017gut, Cai:2017vyk} and generalizations in the form of tensor type models \cite{Witten:2016iux, Gurau:2016lzk, Klebanov:2016xxf, Peng:2016mxj, Ferrari:2017ryl, Itoyama:2017emp, Narayan:2017qtw, Diaz:2017kub, deMelloKoch:2017bvv, Choudhury:2017tax, Prakash:2017hwq, Halmagyi:2017leq, BenGeloun:2017jbi}.
Interesting random matrix theory interpretations have been realized in \cite{You:2016ldz, Garcia-Garcia:2016mno, Cotler:2016fpe, Liu:2016rdi, Krishnan:2016bvg, Garcia-Garcia:2017pzl, Li:2017hdt}.

The original SYK model has a four fermion interaction with a random
coupling which has a gaussian probability distribution with width
$J$. Averaging over the couplings gives rise to a theory with eight
fermi interactions with coupling $J^2$ and an $O(N)$ symmetry, where
$N$ is the number of fermions, making this similar to other vector
models. Like other vector models, this is most easily solved at large
$N$ by making a change of variables to bilocal fields
\cite{jevsak}. For such $O(N)$ models it was proposed in
\cite{Das:2003vw} that these bi-local fields in fact provide a bulk
construction of the dual higher spin theory \cite{Klebanov:2002ja},
with the pair of coordinates in the bi-local combining to provide the
coordinates of the emergent AdS space-time. In $d \geq 2$, the
proposal of \cite{Das:2003vw} was implemented, with additional nonlocal 
transformations on external legs
\cite{Koch:2010cy, Koch:2014mxa, Koch:2014aqa} providing a map between the  bi-local
and  conventional Vasiliev higher spin fields in $AdS_4$. For $d=1$ case (as in the SYK model) 
the simplest identification of the center of mass coordinate
and the relative coordinate of the two points of the bilocal indeed
provides  coordinates of a Poincare patch of {\em lorentzian}
$AdS_2$. In \cite{Jevicki:2016bwu,Jevicki:2016ito} the collective
field theory of the bilocals was developed, providing a transparent
way of obtaining both the bilocal propagator and interactions as well
as the Schwarzian theory of the low energy mode.

The precise bulk dual of the SYK model and its $q$ fermion coupling generalization are
still not well understood. It has been conjectured in \cite{Jensen:2016pah, Maldacena:2016upp, Engelsoy:2016xyb,Forste:2017kwy} that the gravity sector of this model is the Jackiw-Teitelboim model \cite{Teitelboim:1983ux, jackiw} of dilaton-gravity with a negative cosmological constant, studied in \cite{Almheiri:2014cka}, while \cite{Mandal:2017thl} provides strong evidence that it is actually Liouville theory.
(See also \cite{Cvetic:2016eiv, Taylor:2017dly, Diaz:2016kkn, Mezei:2017kmw}).

The spectrum of the SYK model is highly nontrivial. The matter sector of these theories contains an infinite tower of
particles \cite{Polchinski:2016xgd, Maldacena:2016hyu,
  Jevicki:2016bwu}. This is clear from the quadratic action for the
bilocal fluctuations and the resulting bilocal propagator. The kinetic
term in the action contains {\em all powers of the $AdS_2$ laplacian}
which gives rise to rather complicated form of the residues at the
poles of the propagator.  The couplings of these particles are
likewise very complicated, as is clear from the higher point functions
computed in \cite{Gross:2017hcz,Gross:2017aos}.

In \cite{Das:2017pif} it was shown for the $q=4$ model that the exact
spectrum and the bilocal propagator follows from a three dimensional
model. In this 3D realization (where the additional third dimension is used to parametrize the spectrum as in the KK scheme and HS theories), a scalar field with a conventional kinetic
energy term is defined on  $AdS_2 \times I$, where $I = S^1/Z_2$ is a finite
interval with a suitable size. The mass of the scalar field is at the
Breitenlohner-Freedman bound \cite{Breitenlohner:1982bm} of
AdS$_2$. The scalar field satisfies Dirichlet boundary conditions at
the ends and feels an external delta function potential at the middle
of the interval. However, as we will see below, the odd parity modes do not play any role in our construction. This means that one can consider half of the interval
with Dirichlet condition at one end, and a nontrivial boundary
condition determining the derivative of the field at the other end
\footnote{We thank Edward Witten for a clarification on this point.}.
The background can be thought of as coming from the near-horizon
geometry of an extremal charged black hole which reduces the gravity
sector to Jackiw-Teitelboim model with the metric in the third
direction becoming the dilaton of the latter model
\cite{Maldacena:2016upp}. The strong coupling limit of the SYK model
corresponds to a trivial metric in the third direction, while at
finite coupling this acquires a dependence on the AdS$_2$ spatial
coordinate.  At strong coupling a Horava-Witten compactification then
leads to a spectrum of masses in $AdS_2$ which is in exact agreement
with the SYK spectrum. More significantly, a {\em non-standard}
propagator with the end points at the location of the delta function
exactly reproduces the SYK bilocal propagator, if the two coordinates
in the Poincare patch $AdS_2$ are identified with the center of mass
and the relative coordinate of the two points of the bilocal. The nontrivial factors which appear in the SYK propagator from residues at the poles now appear as the values of the wave function at the center of $I$.

As expected, this strong coupling propagator is divergent due to the
divergent contribution of a mode which can be identified with a
reparametrization invariance zero mode in the SYK model. At finite
coupling the zero mode is lifted and gives rise to an ``enhanced
contribution'' proportional to $J$. In the three dimensional model \cite{Das:2017pif} we
adopted the proposal of  \cite{Maldacena:2016upp,
Engelsoy:2016xyb}, and show that to order $1/J$, the poles of the
propagator shift in a manner consistent with the explicit results in
\cite{Maldacena:2016hyu} and the enhanced propagator is reproduced as
well. 

In this paper we show that such a three dimensional picture holds for
generalizations of the SYK model with arbitrary $q$. As shown below, the three dimensional metric on which the scalar lives is now conformal to $AdS_2 \times I$. The scalar field is subject to a non-trivial potential in addition to a delta function at the center of the interval. This reproduces the spectrum exactly. Furthermore, the three dimensional propagator whose points lie at the center of $I$ reproduce the arbitrary $q$ SYK propagator up to a factor which depends only on $q$. We also discuss the large $q$ limit in this picture. In the SYK model the spectrum becomes evenly spaced in this limit. However only one mode - the zero mode - contributes to the propagator since the residues at the other poles vanish. In the three dimensional picture, the different modes appear as KK modes. However in the large $q$ limit we find that the normalized wave function at the center of the interval is nonzero for only one of these modes, in a way consistent with the SYK result.

In section (\ref{sec:qfermion}) we describe the bilocal collective formulation of the model for arbitrary $q$. In section (\ref{sec:3dmodel}) we describe our three dimensional model. Section (\ref{sec:compare}) contains the comparison of the propagator of the three dimensional model with the SYK bilocal propagator. In section (\ref{sec:largeq}) we comment on the large $q$ limit. Section (\ref{sec:conclude}) contains some concluding remarks.

\section{$q$ Fermion SYK Model}
\label{sec:qfermion}
The model is defined by a hamiltonian, with any even $q$,
\begin{equation}
		H \, = (i)^{\frac{q}{2}}\sum_{1\leq i_1 < i_2 <\cdots <
                i_q \leq N}j_{i_1i_2\cdots i_q} \, \chi_{i_1} \,
                \chi_{i_2} \, \cdots \chi_{i_q} \, ,
	\label{Hamiltonian}
	\end{equation}
where $\chi_i$ are Majorana fermions, which satisfy $\{ \chi_i, \chi_j
\} = \delta_{ij}$. The random coupling has a gaussian distribution with 
\ben
<j^2_{i_1i_2\cdots i_q}> = \frac{J^2 (q-1) !}{N^{q-1}}
\label{1-2}
\een
One way to perform the averaging is to use the replica trick.
One does not expect a spin glass
state in this model \cite{Sachdev:2015efa} so that we can restrict to
the replica diagonal subspace \cite{Jevicki:2016bwu}.
At large $N$ this model is efficiently solved by re-writing the theory
in terms of replica diagonal bilocal collective fields \cite{Jevicki:2016bwu, Jevicki:2016ito}.
\begin{equation}
		\Psi(t_1, t_2) \, \equiv \, \frac{1}{N} \sum_{i=1}^N \chi_i(t_1) \chi_i(t_2) \, ,
	\end{equation}
where we have suppressed the replica index.
The corresponding path-integral is
	\begin{equation}
		Z \, = \, \int \prod_{t_1, t_2} \mathcal{D}\Psi(t_1, t_2) \ \mu(\Psi) \, e^{-S_{\rm col}[\Psi]} \, , 
	\label{eq:collective partition function}
	\end{equation}
where $S_{\rm col}$ is the collective action:
	\begin{equation}
		S_{\rm col}[\Psi] \, = \, \frac{N}{2} \int dt \, \Big[ \partial_t \Psi(t, t')\Big]_{t' = t} \, + \, \frac{N}{2} \, {\rm Tr} \log \Psi \, - \, \frac{J^2N}{2q} \int dt_1 dt_2 \, \Psi^q(t_1, t_2) \, .
	\label{S_col}
	\end{equation}
Here the second term comes from a Jacobian factor due to the change of path-integral variable, and the trace is taken over the bi-local time.
One also has an appropriate order $\mathcal{O}(N^0)$ measure $\mu$.
This action being of order $N$ gives a systematic $G=1/N$ expansion, while the measure $\mu$ found as in \cite{Jevicki:2014mfa} begins to contribute at one-loop level (in $1/N$).
As is well known, in the IR, i.e. at strong coupling the kinetic term
can be ignored. There is now an emergent reparametrization
invariance. In this limit the saddle point equation which follow from
$S_{\rm col}[\Psi]$ has the solution
\ben
	\Psi_0(t_1, t_2) \, = \, \frac{b}{|t_{12}|^{\frac{2}{q}}}{\rm
          sgn}(t_{12}) ~~~~~~~b^q = \frac{\tan (\frac{\pi}{q})}{J^2
          \pi} \left(\frac{1}{2} - \frac{1}{q}\right)
\label{1-3}
\een
where we defined $t_{ij}\equiv t_i -t_j$. 

In the following it will be
useful to use the center of mass and relative coordinates
\ben
	t \, = \, \frac{1}{2} \, (t_1 + t_2) \, , \qquad z \,
                = \, \frac{1}{2} \, (t_1 - t_2) \, ,
\label{1-4}
\een
The conformal transformations on $t_1,t_2$ then give rise to
transformations on $t,z$ which are identical to the isometries of
$AdS_2$ with a metric $ds^2 = \frac{1}{z^2} [ -dt^2 +
  dz^2]$. \footnote{Note that this could very well be $dS_2$
  \cite{Maldacena:2016hyu}} Fluctuations around this critical IR
solution, $\Psi_0(t,z)$ defined by
\ben
\Psi(t_1, t_2) \, = \, \Psi_0(t_1, t_2) \, + \, \sqrt{\frac{2}{N}}
 \eta (t,z)
\label{1-5}
\een
can be expanded as
	\begin{equation}
		\eta(t,z) = \int \frac{d\omega}{2\pi} \int \frac{d\nu}{N_\nu} \tilde{\Phi}_{\nu, \omega} u_{\nu, \omega}(t, z)
\label{1-6}
	\end{equation}
where 
\begin{equation}
		u_{\nu, \omega}(t, z) \, = \, {\rm sgn}(z) \, e^{i \omega t} \, Z_{\nu}(|\omega z|) 
\label{1-7}
	\end{equation}
with $Z_{\nu}$ are a complete set of modes which diagonalizes the quadratic kernel \cite{Polchinski:2016xgd},
\begin{equation}
		Z_{\nu}(x) \, = \, J_{\nu}(x) \, + \, \xi_{\nu} \, J_{-\nu}(x) \, , \qquad \xi_{\nu} \, = \, \frac{\tan(\pi \nu/2)+1}{\tan(\pi \nu/2)-1} \, ,
	\label{Znu}.
	\end{equation}
Their normalization and completeness relations are given by
\bea
		\int_0^{\infty} \frac{dx}{x} \, Z^*_{\nu}(x) \, Z_{\nu'}(x) \, & = & \, N_{\nu} \, \delta(\nu-\nu') \nonumber \\
\int \frac{d\nu}{N_{\nu}} \, Z_{\nu}^*(|x|) \, Z_{\nu}(|x'|) \, & = & \, x \, \delta(x-x')
	\label{orthogonality}
\eea
where the normalization factor $N_{\nu}$ is 
\begin{align}
		N_{\nu} \, = \,
		\begin{cases}
			(2\nu)^{-1}  &{\rm for}\ \nu=3/2+2n \\
			2\nu^{-1}\sin\pi\nu \quad &{\rm for}\ \nu=ir \, ,
		\end{cases}
	\label{N_nu}
	\end{align}
In all of the above expressions the integral over $\nu$ is a shorthand for an integral over the imaginary axis and a sum over the discrete values $ \nu=3/2+2n$. The necessity of both the continuous and the discrete spectrum follows from $SL(2,R)$ representation theory \cite{Kitaev:2017hnr}.

The quadratic part of the fluctuation action then becomes
\ben
S_{(2)} \propto J \int d\nu \int d\omega \tilde{\Phi}^\star_{\nu, \omega} \left[
\tilde{k}_c(\nu,q) -1 \right] \tilde{\Phi}_{\nu, \omega}
\label{1-9}
\een
where
\ben
\tilde{k}_c (\nu,q) = \frac{1}{k_c (h,q)}~~~~~~h = \frac{1}{2} + \nu
\label{1-10}
\een
and $k_c(h,q)$ is the eigenvalue of the bilocal kernel derived in \cite{Maldacena:2016hyu}, 
\begin{equation}
k_c(h,q)=-(q-1)\frac{\Gamma\left(\frac{3}{2}-\frac{1}{q}\right)\Gamma\left(1-\frac{1}{q}\right)\Gamma\left(\frac{h}{2}+\frac{1}{q}\right)\Gamma\left(\frac{1}{2}+\frac{1}{q}-\frac{h}{2}\right)}{\Gamma\left(\frac{1}{2}+\frac{1}{q}\right)\Gamma\left(\frac{1}{q}\right)\Gamma\left(\frac{3}{2}-\frac{1}{q}-\frac{h}{2}\right)\Gamma\left(1-\frac{1}{q}+\frac{h}{2}\right)}
\label{kc}
\end{equation}
The spectrum is then given by solving $k_c(h,q) = 1$. Note that $p_m =3/2$ is an exact solution for all $q$.

The bilocal propagator in $(t,z)$ space can be now derived following the same steps as in \cite{Jevicki:2016bwu,Das:2017pif} by substituting the expansion (\ref{1-6}) and using the $(\nu,\omega)$ space propagator which follows from (\ref{1-9}),
\begin{equation}
\mathcal{G}(t,z;t^\prime,z^\prime) \sim -\frac{1}{J} |zz^\prime|^{\frac{1}{2}}
\sum_m\int_{-\infty}^{\infty}\frac{d\omega}{2\pi}e^{-i\omega(t-t^\prime)}\int \frac{d\nu}{N_\nu}\frac{Z^*_\nu(\vert\omega z\vert)Z_\nu(\vert\omega z^\prime\vert)}{\nu^2-p_m^2} (2p_m) R(p_m)
\label{sykprop}
\end{equation}
where $p_m$ denote the solutions of the spectral equation $k_c (p_m +\frac{1}{2},q) = 1$. The factor $R (p_m)$ is the residue of the propagator at the poles $\nu = p_m$,
\ben
R(p_m)  =  \frac{1}{\left(\frac{\partial\tilde k_c(\nu,q)}{\partial\nu}\right)_{\nu=p_m}}
\label{1-11}
\een
and
\begin{equation}
\frac{\partial\tilde k_c(\nu,q)}{\partial\nu}=N_h\left[H_{-1+\frac{h}{2}+\frac{1}{q}}+H_{\frac{1}{2}-\frac{h}{2}-\frac{1}{q}}-H_{\frac{h}{2}-\frac{1}{q}}-H_{-\frac{1}{2}-\frac{h}{2}+\frac{1}{q}}\right]
\label{1-12}
\end{equation}
where $H_n$ denotes the Harmonic number, and 
\begin{equation}
N_h=\frac{\left(\sin\pi h+\sin\frac{2\pi}{q}\right)\Gamma\left(\frac{2}{q}\right)\Gamma\left(2-h-\frac{2}{q}\right)\Gamma\left(1+h-\frac{2}{q}\right)}{\pi q\Gamma\left(3-\frac{2}{q}\right)}
\label{1-13}
\end{equation}
The symbol $\int d\nu$ is as usual a shorthand notation for an integral over the imaginary axis and a sum over discrete values $\nu = \frac{3}{2} + 2n$.
As in the $q=4$ case, when one performs the $\nu$ integral over the imaginary axis there are two sets of poles, the ones at $\nu = \pm p_m$ and at $\nu = \frac{3}{2} + 2n$. The contribution from those latter poles exactly cancel the contribution from the discrete values , and one is finally left with an expression
\ben
\mathcal{G}(t,z;t^\prime,z^\prime) \sim -\frac{1}{J} |zz^\prime|^{\frac{1}{2}}
\sum_m\int_{-\infty}^{\infty}\frac{d\omega}{2\pi}e^{-i\omega(t-t^\prime)}\frac{Z_{-p_m}(\vert\omega\vert z^>)J_{p_m}(\vert\omega\vert z^<)}{N_{p_m}} R_{p_m}
\label{1-14}
\een
where $z^< (z^>)$ is the smaller (larger) of $z,z^\prime$.

As expected, the expression (\ref{1-14}) is divergent since this is a strong coupling proagator. This comes from the mode at $p_m = 3/2$ which is a solution for all $q$. At this value $Z_{-3/2}$ diverges because $\xi_{-3/2}$ diverges. For finite $J$ this mode is corrected by a term which is $O(1/J)$ and this leads to a contribution to the propagator which is $O(J)$ compared to the contribution from the other solutions of $k_c(p_m+1/2,q) =1$.

\section{The Three Dimensional Model}
\label{sec:3dmodel}

We will now write down a model which reproduces the above spectrum exactly and the above propagator up to a function of $q$. The model is that of a single scalar field $\Phi$ with an action
\begin{equation}
\frac{1}{2}\int dt dz dx \sqrt{-g}\left[-g^{\mu\nu}\partial_\mu\Phi\partial_\nu\Phi-V(x)\Phi^2\right]
\label{2-1}
\end{equation}
where the background metric is given by
\begin{equation}
ds^2 = |x|^{\frac{4}{q}-1} \left[ \, \frac{-dt^2 + dz^2}{z^2} + \frac{dx^2}{4|x|(1-|x|)} \right]
\label{2-2}
\end{equation}
and the direction $x$ lies in the interval $-1<x<1$. The space-time is then {\em conformal} to $AdS_2 \times S^1/Z_2$. The potential which appears in (\ref{2-1}) is given by
\begin{equation}
V(x) = \frac{1}{|x|^{\frac{4}{q}-1}}\left[4\left(\frac{1}{q}-\frac{1}{4}\right)^2+ m_0^2+\frac{2V}{J(x)}\left(1-\frac{2}{q}\right) \delta (x) \right]
\label{2-3}
\end{equation}
where $V$ is a constant to be determined below and
\begin{equation}
J(x)=\frac{|x|^{\frac{2}{q}-1}}{2\sqrt{1-|x|}}
\label{2-4}
\end{equation}
The action can be now re-written as 
\bea
S=\frac{1}{2}\int dtdzdx & J(x) & \bigg[(\partial_t\Phi)^2-(\partial_z\Phi)^2-\frac{m_0^2}{z^2}\Phi^2- \\ \nonumber 
& & \frac{4}{z^2} \bigg\{|x| (1-|x|)(\partial_x\Phi)^2+\left(\frac{1}{q}-\frac{1}{4}\right)^2\Phi^2 +\left(1-\frac{2}{q}\right) \frac{V}{2 J(x)} \delta (x) \Phi^2 \bigg\} \bigg]
\label{2-5a}
\eea
We will impose Dirichlet boundary conditions at $x = \pm 1$,
\ben
\Phi (t,z,\pm 1) = 0
\label{2-5}
\een
while the delta function discontinuity in the potential determines the discontinuity at $x=0$ to be
\begin{equation}
{\rm{Lim}}_{\epsilon \rightarrow 0} \left[ |x|^{2/q}\sqrt{1-|x|}\partial_x \Phi \right]^\epsilon_{-\epsilon} = \left(1-\frac{2}{q}\right) V \Phi (t,z,0)
\label{2-6}
\end{equation}
In the following we will be interested in fields which are even under $x \rightarrow -x$. For such fields (\ref{2-6}) implies
\ben
\left[ x^{2/q}\partial_x \Phi \right]_{x=0} = (1-\frac{2}{q}) \frac{V}{2} \Phi(t,z,0)
\label{2-7}
\een
Once we impose this we can restrict to  $ 0 < x < 1$ and forget about  the delta function.
This is what we will do in the rest of the paper.

Performing an integration by parts and ignoring the boundary term the action becomes
\begin{equation}
S = \frac{1}{2} \int_{-\infty}^\infty dt \int_0^\infty dz \int_0^1 dx~J(x)~\Phi {\cal{D}}_0 \Phi
\label{2-8}
\end{equation}
where
\begin{equation}
\mathcal{D}_0 = -\partial_t^2+\partial_z^2-\frac{m_0^2}{z^2}+\frac{4}{z^2}\left[x(1-x)\partial_x^2+\left[\frac{2}{q}-x\left(\frac{1}{2}+\frac{2}{q}\right)\right]\partial_x-\left(\frac{1}{q}-\frac{1}{4}\right)^2\right]
\label{2-9}
\end{equation}
This operator is hermitian with the measure $dx J(x)$.

\subsection{The Spectrum}

To diagonalize $\mathcal{D}_0$ we first find solve the eigenvalue problem for the operator inside the square bracket in (\ref{2-9}) in the domain $0 < x < 1$,
\begin{equation}
\left[x(1-x)\partial_x^2+\left[\frac{2}{q}-x\left(\frac{1}{2}+\frac{2}{q}\right)\right]\partial_x-\left(\frac{1}{q}-\frac{1}{4}\right)^2\right]\phi_k(x)=-\frac{k^2}{4}\phi_k(x)
\label{2-10}
\end{equation}
The general solution of this equation is 
\begin{equation}
\phi_k(x)=A~_2F_1(a,b;c,x)+x^{1-c}B~_2F_1(a-c+1,b-c+1;2-c;x)
\label{2-11}
\end{equation}
where $_2F_1$ denotes the usual Hypergeometric function and
\begin{equation}
a  = \frac{1}{q}-\frac{1}{4}-\frac{k}{2}~~~~~~
b  =  \frac{1}{q}-\frac{1}{4}+\frac{k}{2}~~~~~~
c  =  \frac{2}{q}
\label{abc}
\end{equation}
Imposing the boundary condition (\ref{2-7}) gives
\ben
B=\frac{V}{2}A
\label{2-12}
\een
while imposing (\ref{2-5}) gives
\begin{equation}
A_2F_1(a,b;c,1)+B_2F_1(a-c+1,b-c+1;2-c;1)=0
\label{2-13}
\end{equation}
Using
\begin{equation}
_2F_1(a,b;c,1)=\frac{\Gamma(c)\Gamma(c-a-b)}{\Gamma(c-a)\Gamma(c-b)}
\label{2-14}
\end{equation}
and combining (\ref{2-13}) and (\ref{2-12}) we get
\begin{equation}
\frac{\Gamma\left(\frac{5}{4}-\frac{1}{q}-\frac{k}{2}\right)\Gamma\left(\frac{2}{q}\right)\Gamma\left(\frac{5}{4}-\frac{1}{q}+\frac{k}{2}\right)}{\Gamma\left(2-\frac{2}{q}\right)\Gamma\left(\frac{1}{4}+\frac{1}{q}-\frac{k}{2}\right)\Gamma\left(\frac{1}{4}+\frac{1}{q}+\frac{k}{2}\right)}=-\frac{V}{2}
\label{2-15}
\end{equation}
where we have used the values of $a,b,c$ in (\ref{abc}).

Remarkably if we choose
\begin{equation}
V=2(q-1)\frac{\Gamma\left(\frac{3}{2}-\frac{1}{q}\right)\Gamma\left(1-\frac{1}{q}\right)\Gamma\left(\frac{2}{q}\right)}{\Gamma\left(\frac{1}{2}+\frac{1}{q}\right)\Gamma\left(2-\frac{2}{q}\right)\Gamma\left(\frac{1}{q}\right)}
\label{2-16}
\end{equation}
and define $h = k + 1/2$ the condition (\ref{2-15}) becomes
\begin{equation}
k_c(h,q)=1
\label{2-17}
\end{equation}
where $k_c(h,q)$ is precisely the SYK spectrum for arbitrary $q$ given by (\ref{kc}).
The significant point of course is that $V$ given by (\ref{2-16}) depends only on $q$.

\subsection{The two point function}

Using the eigenfunctions in the previous subsection we can now expand the three dimensional field in terms of a complete basis as follows
\begin{equation}
\Phi(t,z,x) = \int \frac{dkd\nu d\omega}{N_\nu}e^{-i\omega t} \vert z\vert^{1/2}Z_\nu(\vert\omega z\vert)\varphi_k(x)~\chi ( \omega, \nu, k )
\label{2-18}
\end{equation}
where the combinations of Bessel functions $Z_\nu$ have been defined in (\ref{Znu}) and $\varphi_k (x)$ are 
\ben
\varphi_k(x) = _2F_1(a,b;c,x)+x^{1-c} \frac{V}{2}~ _2F_1(a-c+1,b-c+1;2-c;x)
\label{2-19}
\een
where $a,b,c$ are given in (\ref{abc}) and $V$ is given in (\ref{2-16}). The functions $\varphi_k(x)$ are orthogonal with the measure factor $J(x)$
\begin{equation}
\int_0^1 dx J(x) \varphi_k(x)\varphi_{k^\prime}(x)=C_1(k)\delta_{k,k^\prime}
\label{2-20}
\end{equation}
The action now becomes
\begin{equation}
S=\frac{1}{2}\int \frac{dk d\nu d\omega}{N_\nu} C_1(k)(\nu^2-\nu_0^2)\chi (\omega,\nu, k) \chi (-\omega,\nu, k )
\label{2-21}
\end{equation}
where
\begin{equation}
\nu_0^2 = k^2+m_0^2+\frac{1}{4}
\label{2-22}
\end{equation}
Let us now choose
\begin{equation}
m_0^2 = -1/4
\end{equation}
so that one finally has $\nu_0=k$. The two point function of $\chi ( \omega\nu k )$ is then given by
\begin{equation}
\langle\chi(\omega,\nu,k)\chi(-\omega,\nu,k)\rangle =\frac{N_\nu}{C_1(k)(\nu^2-k^2)}
\label{2-23}
\end{equation}
The position space 3d propagator is then given by
\begin{equation}
\langle\Phi(t,z,x)\Phi(t^\prime,z^\prime,x^\prime)\rangle = \vert zz^\prime\vert^{\frac{1}{2}}\sum_m C(p_m,x,x^\prime)\int\frac{d\omega}{2\pi}e^{-i\omega(t-t^\prime)}\int\frac{d\nu}{N_\nu}\frac{Z^\star_\nu(\vert\omega z\vert)Z_\nu(\vert\omega z^\prime\vert)}{\nu^2-p_m^2}
\label{3dprop}
\end{equation}
where
\begin{equation}
C(p_m,x,x^\prime)=\frac{\varphi_{p_m}(x)\varphi_{p_m}(x^\prime)}{C_1(p_m)}
\label{cpm}
\end{equation}
This can be regarded as a sum of $AdS_2$ propagators. However it is important to note that these are non-standard propagators.

\section{Comparison of the 3d and SYK propagator}
\label{sec:compare}

We now show that the propagator (\ref{3dprop}) evaluated at $x = x^\prime = 0$ agrees with the SYK propagator (\ref{sykprop}) up to an overall factor which depends on $q$. The values of $p_m$ over which the two expressions need to be summed have been already seen to be identical, so we need to compare the coefficients which appear. To compare (\ref{sykprop}) and (\ref{3dprop}) we need to compute the quantity
\begin{equation}
\frac{C(p_m,0,0)}{2 p_m R (p_m)}
\label{2-25}
\end{equation}
and show that this is independent of $p_m$. Here
\begin{equation}
C_1(p_m) = \int_0^1 dx J(x) \varphi_{p_m}(x)\varphi_{p_m}(x)
\label{2-26}
\end{equation}
We have not been able to evaluate this integral analytically, but have performed this numerically to high precision. 
In Table (\ref{table1}) we tabulate the values of the relevant quantities for various values of $q$ and $p_m$ which solve the spectrum equation, and compare them with the corresponding factors which appear in (\ref{sykprop})

\begin{table}[h!]
\begin{center}
\begin{tabular}{| c | c | c | c | c |} 
\hline
$q$ & $p_m$ & $C(p_m,0,0)$ & $2p_m R(p_m)$ & $\frac{C(p_m,0,0)}{2p_m R(p_m)}$ \\ \hline
6 & 1.5 &  0.415724 & 1.09987 & 0.377976 \\
& 3.07763 & 0.566693 & 1.49928 & 0.377976 \\
& 4.95427 & 0.474406 & 1.25512 & 0.377976 \\
& 6.90849 & 0.409177 & 1.08255 & 0.377976 \\
& 8.88613 & 0.366967 & 0.970874 & 0.377976 \\ \hline
8 & 1.5 & 0.344227 & 1.27788 & 0.269374 \\
& 2.95416 & 0.357211 & 1.32608 & 0.269374 \\
& 4.835 & 0.241026 & 0.894764 & 0.269374 \\
& 6.79849 & 0.188249 & 0.698838 & 0.269374 \\
& 8.78225 & 0.159144 & 0.590793 & 0.269374 \\ \hline
12 & 1.5 & 0.256089 & 1.47882 & 0.173171 \\
& 2.81505 & 0.175464 & 1.01324 &  0.173171 \\
& 4.71763 & 0.0925242 & 0.534294 &  0.173171 \\
& 6.69343 & 0.0660067 & 0.381165 &  0.173171 \\
& 8.68356 & 0.0529405 & 0.305712 &  0.173171 \\ \hline
20 & 1.5 & 0.169337 & 1.66312 & 0.101819 \\
& 2.69405 & 0.0678495 & 0.666375 & 0.101819 \\
& 4.62734 & 0.0287883 & 0.28274 & 0.101819 \\
& 6.61348 & 0.0192516 & 0.189077 & 0.101819 \\
& 8.60817 & 0.0148617 & 0.145963 & 0.101819 \\ \hline
50 & 1.5 & 0.0745898 & 1.85438 & 0.0402235 \\
& 2.57914 & 0.0115317 & 0.286691 & 0.0402235 \\
& 4.54971 & 0.00396415 & 0.0985529 & 0.0402235 \\
& 6.54453 & 0.00251532 & 0.0625335 &  0.0402235 \\ \hline
\end{tabular}
\end{center}
\caption{ Comparison of Factors Appearing in the 3d and SYK Propagators }
\label{table1}
\end{table}

For a given value of $q$ the value of the ratio (\ref{2-25}) is independent of $p_m$ upto 13 decimal places. This shows that this ratio is only a function of $q$ which we denote by $f(q)$. 
These results show that for any given $q$, the non-standard propagator of the 3d model with the two points at $x=0$ is proportional to the SYK propagator. The data also shows that $f(q)$ decreases with $q$. 

As for $q=4$, the propagator (\ref{3dprop}) is actually divergent from the contribution of the $p_m =3/2$ mode, as expected from the SYK propagator at infinite coupling. We expect that a modification of the three dimensional background would reproduce the enhanced propagator of this mode, as happened for $q=4$ \cite{Das:2017pif}.

\section{The large $q$ limit}
\label{sec:largeq}

In the $q \rightarrow \infty$ limit, $p_m =3/2$ remains a solution, while the other solutions of the spectral equation (\ref{kc}) become very simple,
\ben
p_m = 2m + \frac{1}{2} + \frac{2}{q} \frac{2m^2 +m +1}{2m^2 + m -1} + \cdots~~~~~~~m = 1,2,\cdots
\label{3-1}
\een
To calculate the contribution to these poles to the SYK propagator consider the residue $R(p_m)$ in (\ref{1-11}). In a $1/q$ expansion we find that for $p_m =3/2$
\ben
R(3/2) = \frac{2}{3} - \frac{1}{q} \left( \frac{5}{2} + \frac{\pi^2}{3} \right) + O(1/q^2)
\label{3-2}
\een
while for the other solutions we get \footnote{We thank Pranjal Nayak for this calculation.}
\ben
R(p_m) \rightarrow \frac{1}{q} \frac{4(2m^2+m)}{(2m^2+m-1)^2} + O(1/q^2)
\label{3-3}
\een
Thus only the pole at $p_m =3/2$ has a non-vanishing residue in the large $q$ limit. Of course the strong coupling propagator is infinite from contribution of the $p_m =3/2$ mode. However a finite $J$ correction would lead to a nonzero contribution proportional to $J$ \cite{Maldacena:2016hyu}.

In the three dimensional picture this happens because of the different large $q$ behavior of the wave function at $x=0$ for $p_m =3/2$ compared to the other values of $p_m$.
For large enough $q$ we can use $p_m = 2m+1$  as a good approximation to the solution of the spectral equation.
In Table (\ref{large-q}) we tabulate the values of the square of the wave function at $x=x'=0$, i.e. $C(p_m,0,0)$, the value of the quantity $2p_m R(p_m)$ which appears in the SYK propagator and the quantity 
$q f(q)$ where 
\ben
f(q) = \frac{C(p_m,0,0)}{2p_m R(p_m)}
\label{3-5}
\een
for large values of $q$, for different values of $m$.  We have checked that for the values of $q$ which we have used, $p_m = 2m+1$ is indeed a very good approximation to the exact solution of $k_c(h,q)=1$. We then tabulate this quantity for given $m$ for various values of $q$. The results show that $q f(q)$ is a constant to very high accuracy. Using (\ref{3-2}) and (\ref{3-3})  we then conclude that the wave function at $x=0$ for $p_m \neq 3/2$ modes vanishes as $\frac{1}{q^2}$, while that for the $p_m = 3/2$ this vanishes as $\frac{1}{q}$.

\begin{table}[h!]
\begin{center}
\begin{tabular}{| c | c | c | c | c |} 
\hline
$p_m=2m+1/2$ & $q$ & $C_1(p_m,0,0)$ & $2p_mR(p_m)$ &$\frac{qC(p_m,0,0)}{2p_mR(p_m)}$\\ \hline

3/2 & 500 & 125.908 &1.98465& 2.00093 \\
& 600 & 150.908 &1.9872& 2.00077 \\
& 800 &  200.908   &1.99039& 2.00058 \\ 
&1000 &  250.908 &1.9923& 2.00046 \\   \hline

5/2 & 500 &  74442.2 &0. 003356& 2.00093 \\
& 600 &  107330 &0. 002794& 2.00077 \\
& 800 &    191107 &0. 002092& 2.00058 \\ 
&1000 &  298883 &0. 00167& 2.00046 \\   \hline

9/2 & 500 &  137225 &0. 00182& 2.00093 \\
& 600 &  198038 &0. 00151& 2.00077 \\
& 800 &   352937 &0.001133& 2.00058 \\ 
&1000 &   552280 &0. 000905& 2.00046 \\  \hline

21/2 & 500 & 321462  &0.00077& 2.00093 \\
& 600 & 464316  &0.000645& 2.00077 \\
& 800 &  828595  &0.000484& 2.00058 \\ 
&1000 & $1.279\times 10^6 $ &0.000385& 2.00084 \\  \hline

41/2 & 500 &625414   &0. 000399& 2.00093 \\
& 600 & 904123  &0. 000331& 2.00077 \\
& 800 & $1.615\times 10^6$   &0. 000247& 2.00082 \\  
&1000 & $2.52\times 10^6$   &0. 000197& 2.00065 \\  \hline

\end{tabular}
\end{center}
\caption{Large-q behavior of the wave function at $x=x'=0$ for different values of $p_m$}
\label{large-q}
\end{table}

\begin{figure}[h]
\setlength{\abovecaptionskip}{0 pt}
\centering
\includegraphics[scale=0.5]{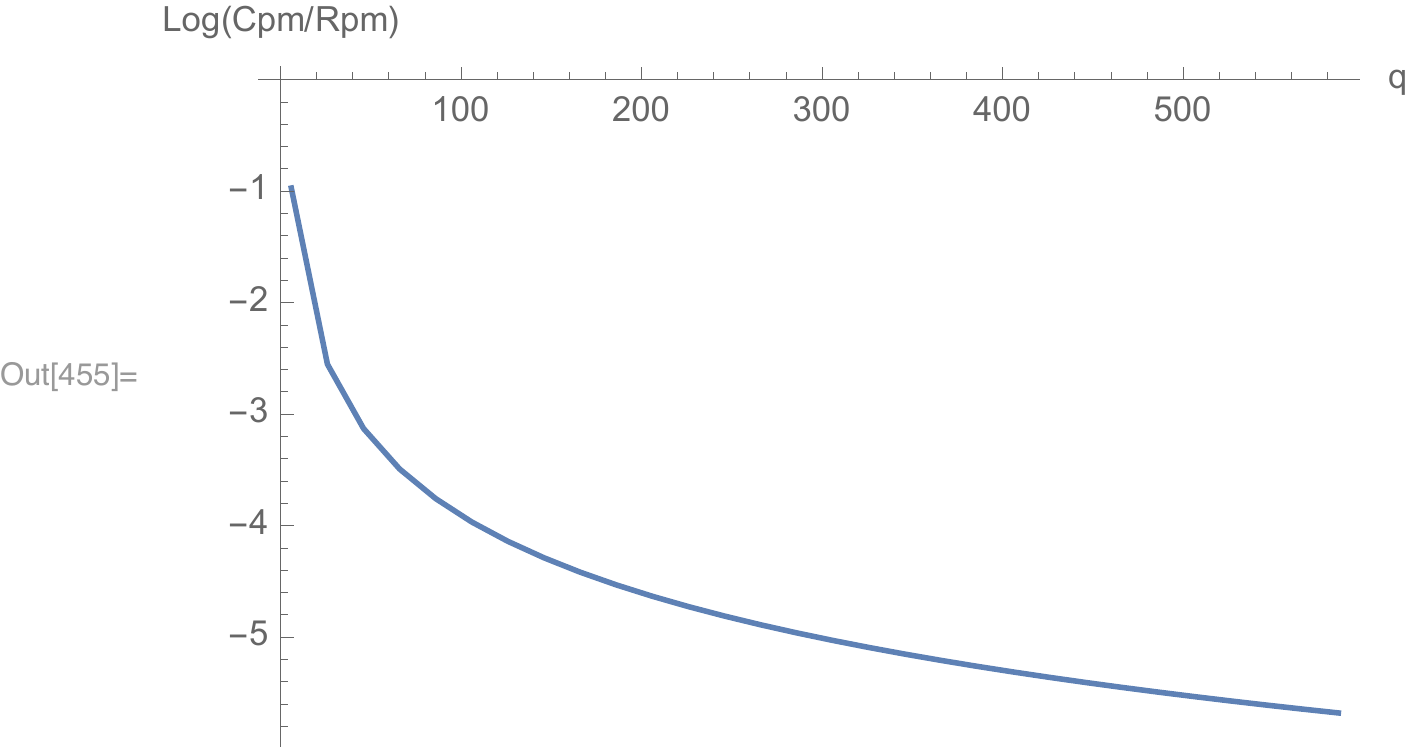}
\caption{Plot of $\log f(q)$ for $p_m=3/2$. For large $q$ the data fits well with the function $f(q) \sim \frac{2}{q}$}
\label{fq}
\end{figure}

A plot of $f(q)$ for $p_m=3/2$ is given in Figure (\ref{fq}).

This behavior provides an understanding of the decoupling of the other modes in the tower at large $q$. Note that the other modes are still present, though they do not contribute to the propagator. In fact the large $q$ limit is subtle. If we perform a $1/q$ expansion of the collective action for the bilocal field, (\ref{S_col}), using the parametrization used in \cite{Maldacena:2016hyu} one ends up with a Liouville theory in the $(t,z)$ space for all values of the suitably rescaled coupling \cite{dgjs3} \footnote{The Dyson-Schwinger equation in the $q \rightarrow \infty$ limit is already known to be akin to the Liouville equation \cite{Maldacena:2016hyu}. Here we are making a stronger statement about the bilocal field itself, not just about its saddle point value.}. This has a conventional kinetic term - so that one seems to get a single two dimensional field, the corresponding pole of the propagator being precisely the $p_m=3/2$ mode. The other modes are simply absent in this treatment, and seem to be recovered due to nonlocal 1/q interactions. 

\section{Conclusions} 
\label{sec:conclude}

It is remarkable that the complicated tower of states which appear in the SYK model can be understood as a KK tower for arbitrary $q$. We want to emphasize that while reproducing the spectrum is already quite interesting, the agreement of the propagator with the bilocal propagator is highly non-trivial. This gives a strong evidence that a three dimensional space-time is an essential ingredient of the full dual to the SYK model. 

Unlike the $q=4$ case we do not yet have a natural understanding of the three dimensional background at arbitrary $q$ in terms of a near horizon geometry of a black hole. We hope to be able to achieve this understanding. That will provide a natural way to understand the finite $J$ correction and in particular the enhanced propagator of the $p_m =3/2$ mode.

In this paper we have not addressed the question of interactions of the bilocals.
These have been considered in \cite{Gross:2017hcz,Gross:2017aos} and are expected to  follow from the cubic and higher order terms in the collective field theory of \cite{Jevicki:2016bwu}. 
It will be interesting to see what kind of interactions in the three dimensional model reproduce these and investigate their locality (or lack thereof) properties 
\footnote{In \cite{Gross:2017aos} a different 3d background is shown to reproduce the large-$q$ spectrum.}.
An important aspect of the 3d picture is that while the propagator can be written as a sum of $AdS_2$ propagators, the latter are {\em non-standard} propagators. While they do vanish at the boundary, they have different boundary conditions at the Poincare horizon. A second unusual aspect is that the space of bilocals always gives rise to Lorentzian $AdS_2$ even if we start out with an euclidean theory. 
The  issues raised above  require a better understanding of the bulk theory , we will address them in a forthcoming 
publication \cite{dgjs2}.

\acknowledgments

We thank Juan Maldacena and Pranjal Nayak for discussions, and Ed Witten for a correspondence. This work of AJ and KS is supported by the Department of Energy under contract DE-SC0010010. The work of KS  
 is also supported by the Galkin Fellowship Award at Brown University. The work of SRD  and AG is partially supported by the National Science Foundation grant NSF-PHY-1521045. S.R.D. would like to thank Yukawa Institute for Theoretical Physics, Kyoto University and Tata Institute of Fundamental Research, Mumbai for hospitality during the completion of this work. AG would like to thank the lecturers and participants of the TASI 2017 program, and the University of Colorado, Boulder for hospitality during the completion of this work.

\end{document}